\documentclass[preprint,12pt]{elsarticle}

\usepackage{amssymb}
\usepackage{xcolor}

\journal{Cell Systems}

\begin{document}

\begin{frontmatter}



\title{Cellular Griffiths-like phase}


\author[unesp1]{Lucas Squillante}
\address[unesp1]{São Paulo State University (Unesp), IGCE - Physics Department, Rio Claro - SP, Brazil}
\author[unesp1]{Isys F. Mello}
\author[iceland]{Luciano S. Ricco}
\affiliation[iceland]{Science Institute, University of Iceland, Dunhagi-3, IS-107, Reykjavik, Iceland}
\author[unesp2]{Marcos F. Minicucci}
\affiliation[unesp2]{Department of Internal Medicine, Botucatu Medical School, UNESP – Univ Estadual Paulista, Botucatu, Brazil}
\author[an1,an2]{Aniekan Magnus Ukpong}
\affiliation[an1]{Theoretical and Computational Condensed Matter and Materials Physics Group, School of Chemistry and Physics, University of KwaZulu-Natal, Pietermaritzburg, South Africa}
\affiliation[an2]{National Institute for Theoretical and Computational Sciences (NITheCS), KwaZulu-Natal, South Africa}
\author[unesp3]{Antonio C. Seridonio}
\address[unesp3]{São Paulo State University (Unesp), Department of Physics and Chemistry, Ilha Solteira - SP, Brazil}
\author[unesp1]{Roberto E. Lagos-Monaco}
\author[unesp1]{Mariano de Souza}
\ead{mariano.souza@unesp.br}

\begin{abstract}
Protein compartmentalization in the frame of a liquid-liquid phase separation is a key mechanism to optimize spatiotemporal control of biological systems. Such a compartmentalization process reduces the intrinsic noise in protein concentration due to stochasticity in gene expression. Employing Flory-Huggins solution theory, Avramov/Casalini's model, and the Gr\"uneisen parameter, we unprecedentedly propose a cellular Griffiths-like phase (CGLP), which can impact its functionality and self-organization. The here-proposed CGLP is key ranging from the understanding of primary organisms' evolution to the treatment of diseases. Our findings pave the way for an alternative Biophysics approach to investigate coacervation processes.
\end{abstract}




\end{frontmatter}



\clearpage

\section{Introduction}

\begin{figure*}[t]
\centering
\includegraphics[width=\textwidth]{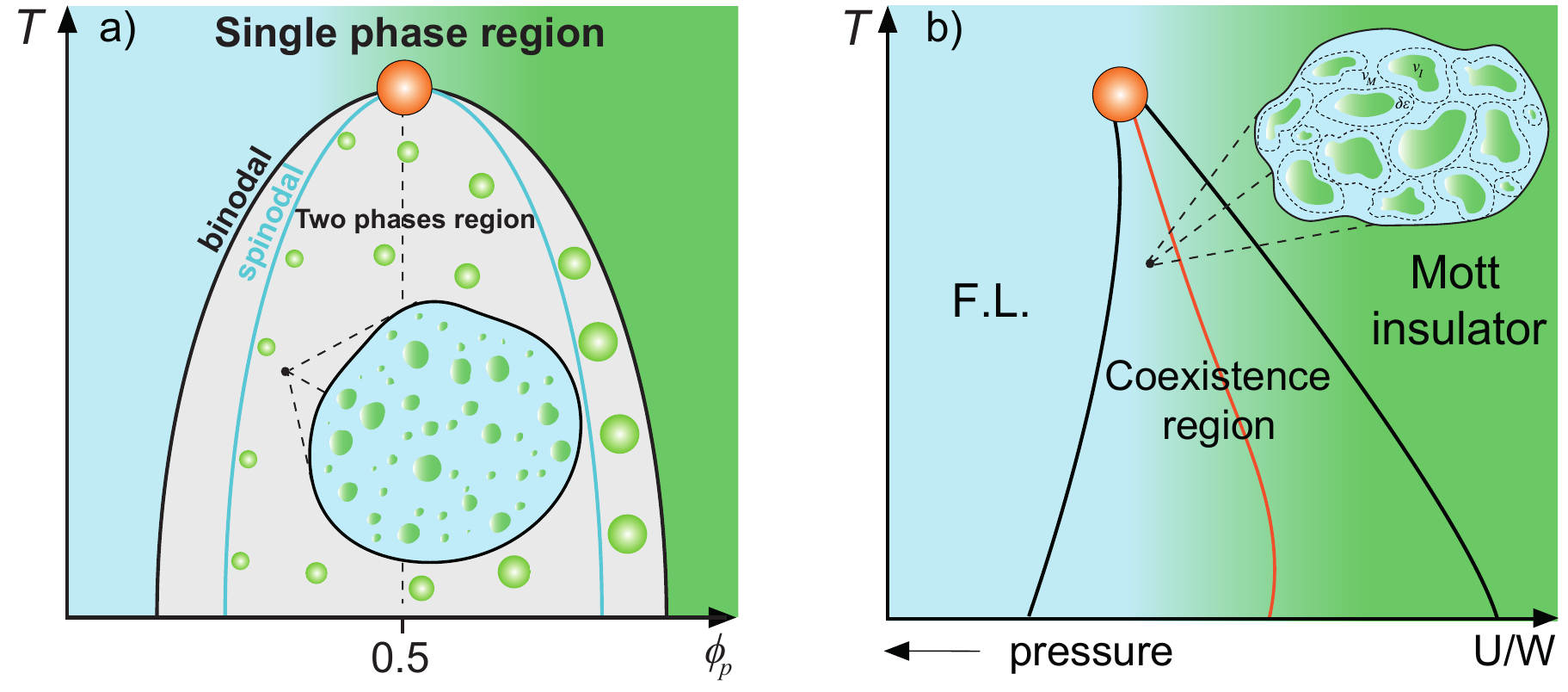}
\caption{\footnotesize a) Schematic representation of the temperature $T$ \emph{versus} protein concentration $\phi_p$  phase diagram depicting the single and the two phases regions separated by the binodal line, which is governed by a critical point (orange color). The spinodal line is also depicted (blue line). In the two-phases region, protein droplets emerge within the cell. A cell containing various protein droplets (green circles) embedded in a solvent (blue color) is schematically depicted zoomed in within the two-phases region. The blue and green color gradient background represents the increasing of proteins inside the cell. b) Schematic representation of the $T$ \emph{versus} $U/W \propto {\wp}^{-1}$ phase diagram of the molecular system $\kappa$-(BEDT-TTF)$_2$Cu$_2$(CN)$_3$ \cite{Mariano2020(3)}, where $U$ is the on-site Coulomb repulsion, $W$ the bandwidth, and ${\wp}$ pressure, showing the coexistence region between Fermi-liquid F.L. (metal) and Mott insulator. The finite-$T$ critical end point is depicted (orange color). The blue and green background gradient represents the metallic and Mott insulating phases, respectively. In the coexistence region, insulating puddles (green) embedded in a metallic matrix (blue background) with their corresponding volumes, namely, $v_M$ and $v_I$, are depicted. The interaction parameter $\delta\varepsilon$ is indicated. Panel b) is adapted from Ref.\,\cite{Mariano2020(3)}.}
\label{Fig-0}
\end{figure*}

It has been known for decades that Physics and Biology are closely linked \cite{Morgan1927}. Indeed, E. Schr\"odinger proposed in his seminal book entitled ``\emph{What is life? The Physical Aspect of the Living Cell}'' \cite{Schrodinger1944} that \emph{the most essential part of a living cell, the chromosome fiber, may suitably be called an aperiodic crystal}, opening a new avenue in molecular Biology \cite{Dronamraju1999}.
Nowadays, the so-called cellular liquid-liquid phase separation (LLPS), i.e., protein compartmentalization [Fig.\,S1a)], is of broad interest since it can be related to disease control, genome stability, and even to the immunity control in plants, cf.\,Refs.\,\cite{Klosin2020,Riback2020, MacMicking2021, Wang2021}. It has been proposed that LLPS is a key mechanism to reduce the noise strength \cite{Banani2017}. A deep understanding of the LLPS dynamics is also relevant for unveiling the possible formation process of primordial organisms in prebiotic Earth, and the consequent evolution to more complex cellular constituents. Making use of an adapted version of Avramov/Casalini's model \cite{Avramov1988,Casalini2004}, usually employed for glassy systems, we investigate the dynamics associated with the LLPS in terms of the Flory-Huggins solution (FHS) theory and the Gr\"uneisen parameter (GP). We present a new approach to investigate cell criticality in terms of the here-proposed cellular Griffiths-like phase, hereafter CGLP, cf.\,Fig.\,\ref{Fig-0}a). It is worth recalling that in the canonical magnetic Griffiths phase either magnetized or non-magnetized rare regions are embedded in a paramagnetic or a ferromagnetic matrix, respectively \cite{Griffiths1969,Vojta2013}. In the present case, we consider the random spatially distributed protein droplets as the rare ferromagnetic regions in a direct analogy to the magnetic Griffiths phase \cite{Griffiths1969}. Hence, the protein droplets (rare regions) can be naturally regarded as colloid-rich while the diluted phase as colloid-poor \cite{Veis2011}, cf.\,Fig.\,\ref{Fig-0}a). The idea of Griffiths-like phases has been recently flourished to the Mott transition with the so-called electronic Griffiths-like phase \cite{Mariano2020(3),Miranda2009,Kanoda2020}, cf.\,Fig.\,\ref{Fig-0}b), which motivated the present work. A Griffiths-like phase was also reported for other biological systems, e.g., brain criticality \cite{Munoz2013,Tragtenberg2016}. It is well-known that the FHS theory was built in a mean-field approximation \cite{Colby2003}, being the Thermodynamic conditions to determine, for instance, the spinodal and binodal lines, as well as the behavior of particular physical quantities in these regions, universal \cite{debenedetti,sethna}. Our proposal is not strict to the FHS theory and can be employed to other models, such as the Voorn-Overbeek \cite{overbeek}, random-phase approximation \cite{rpa}, and the Poisson-Boltzmann cell \cite{stuart}. We only make use of the FHS theory as a \emph{working horse} to showcase an universal behavior of the Gr\"uneisen parameter in the two-phases region, as well as its implications in terms of the LLPS dynamics within living cells.  \newline

\section{Results}

The so-called Gr\"uneisen ratio $\Gamma$ represents the singular contribution to the effective Gr\"uneisen parameter, i.e., the ratio of the isobaric thermal expansivity $\alpha_{\wp}$ to the isobaric heat capacity $c_{\wp}$, being extensively employed as a \emph{smoking-gun} to explore critical phenomena, phase transitions, as well as to quantify caloric effects \cite{Mariano2016, Mariano2020, MarianoPRL,Mariano2019(2),Marianoarxiv, Mariano2020(3), Mariano2021}. The definition of $\Gamma$ reads \cite{Zhu}:
\begin{equation}
\Gamma = \frac{\alpha_{\wp}}{c_{\wp}} = -\frac{1}{V_m T}\frac{\left(\frac{\partial S}{\partial {\wp}}\right)_T}{\left(\frac{\partial S}{\partial T}\right)_{\wp}} = \frac{1}{V_m T}\left(\frac{\partial T}{\partial {\wp}}\right)_S,
\label{gamma}
\end{equation}
where $V_m$, $S$, ${\wp}$, and $T$ are, respectively, the molar volume, entropy, pressure, and temperature.

\begin{figure*}
\centering
 \includegraphics[width=\textwidth]{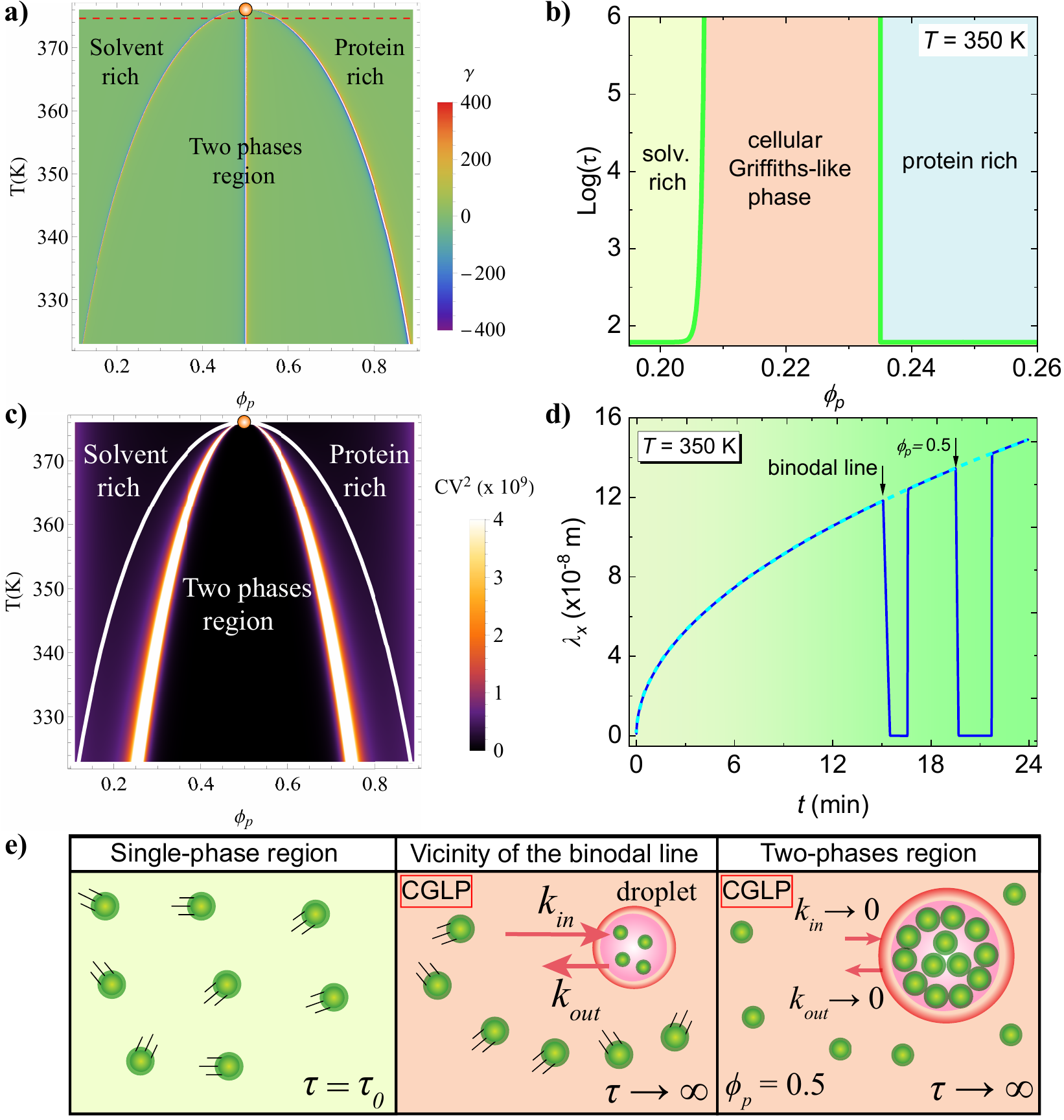}
\caption{\footnotesize a) Density plot of the temperature $T$ \emph{versus} protein concentration $\phi_p$ \emph{versus} effective Gr\"uneisen parameter $\gamma$ depicting both the solvent rich, protein rich, and the two phases region governed by a critical point (orange color). The red dashed line depicts the Ginzburg criteria for the FHS theory considering $N = 2000$. b) Logarithm of the protein diffusion time $\tau$ \emph{versus} $\phi_p$ (green solid line) for $T$ = 350\,K, where the cellular Griffiths-like and both solvent and protein rich phases are depicted. c) Density plot of $T$ \emph{versus} $\phi_p$ \emph{versus} the coefficient of variation $CV^2$ (noise strength). The white solid line is the binodal line. d) Brownian displacement $\lambda_x$ in the $x$ direction \emph{versus} time $t$ for $T$ = 350\,K \cite{SM}. Upon crossing the binodal line and $\phi_p$ = 0.5, $\lambda_x \rightarrow$ 0. The cyan dotted line represents a typical $\lambda_x \propto \sqrt{t}$ behavior. e) Schematic representation of the cellular Griffiths-like phase (CGLP). In the single-phase region, the proteins (green spheres) diffuse in the solvent matrix with a characteristic diffusion time $\tau_0$. As $\phi_p$ is increased and the vicinity of the binodal line is achieved, there is a phase separation and protein droplets emerge and start to grow, so that $k_{in} > k_{out}$. The diffusion time $\tau \rightarrow \infty$ and the CGLP is achieved. Also, at $\phi_p = 0.5$ the CGLP sets in, where $k_{in}$, $k_{out} \rightarrow 0$ and $\tau \rightarrow \infty$. More details in the main text.}
\label{Fig-1}
\end{figure*}

We consider that $V_m = V_{tot}/n$, being $V_{tot}$ the total volume of the cell and $n$ the total number of protein and solvent particles embedded in the cell. Assuming that $V_{tot}$ is fully composed by the sum of protein and solvent particles, $V_{tot}/n$ yields the generic volume $v_p$ occupied by a single protein/solvent particle inside the cell. In our calculations, we make use of $v_p \simeq 10^{-25}$\,m$^3$, which represents the typical volume associated with a single protein, cf.\,Ref.\,\cite{Klosin2020}. Hence, employing the temperature and protein concentration dependences of the free energy of mixture $\Delta F$ \cite{SM}, we compute $\Gamma$ in the frame of the FHS theory on the verge of the binodal line and critical point as a function of the protein concentration $\phi_p$ \cite{SM}. Considering the dimensionless molecular lengths of both protein $N_p$ and solvent $N_s$ \cite{SM}, the Ginzburg criterion must be obeyed in order to determine the region close to the critical point, in which the mean-field character regarding the FHS theory is no longer valid \cite{Colby2003}. Essentially, the Ginzburg criterion for a symmetric polymer mixture is given by $(T-T_c)/T_c \approx 1/N$ \cite{Colby2003}, where $T_c$ is the critical temperature and $N = N_p = N_s$ the molecular length, i.e., the bigger $N$ the closest to the critical point the FHS theory is applicable. Hence, in our analysis we have employed a textbook example regarding a symmetric polymer blend of hydrogenated/deuterated polybutadiene with $N = N_p = N_s = 2000$, being for this case the $T$ range of validity of the FHS theory given by $\Delta T \approx 0.9995\,T_c$, cf.\,horizontal red dashed line in Fig.\,\ref{Fig-1}a). By using such a textbook example, $T_c \approx 100$\,$^{\circ}\mathrm{C}$ and phase separation occurs at a high-temperature, which is clearly deleterious for biological systems. However, we have employed the physical parameters of such polymer blend solely as an example of application of our proposal, which can be adapted to suit the biological problem in hand. Employing $\Delta F$ \cite{SM}, we have:
\begin{eqnarray}
&&\wp(T,\phi_p) = k_B T[N_p -N_s -\chi N_p N_s + 2\chi N_pN_s \phi_p + \nonumber\\
&&+ N_p \log(1-\phi_p)-N_s \log(\phi_p)]\times({N_pN_s V_{tot}})^{-1},
\label{pressure}
\end{eqnarray}
where $k_B$ is Boltzmann constant and $\chi$ the Flory interaction parameter \cite{SM}. In the present case, $\wp$ is the osmotic pressure, i.e., the $\Delta F$ variation upon increasing the total volume of the proteins, which in turn dictates diffusion processes within the cell \cite{Colby2003,SM} . Employing Eq.\,\ref{pressure} and the typical $T$-dependence of $\chi$, namely $\chi \cong A + B/T$ \cite{Colby2003,Banani2017}, where $A$ and $B$ represent, respectively, the entropic and enthalpic contributions to $\chi$ \cite{SM}, $\Gamma$ reads:
\begin{eqnarray}
&&\Gamma(T,\phi_p) = (N_pN_sV_{tot})\times \{k_Bv_p[-BN_pN_s + \nonumber\\
&& + 2BN_pN_s\phi_p + N_pT - N_sT - AN_pN_sT + \nonumber \\
&&+ 2AN_pN_s\phi_pT + N_pT\log{(1 - \phi_p)} - N_sT\log{(\phi_p)}]\}^{-1}.
\label{gammafinal}
\end{eqnarray}
In the vicinity of the binodal line, $\Gamma$ is enhanced and changes sign \cite{SM} indicating that phase separation takes place and protein droplets begin to be formed. Also, for $\phi_p$ = 0.5, $\Gamma$ presents a divergent-like behavior and changes sign, being such a feature in $\Gamma$ reminiscent of a first-order-type phase transition \cite{Mariano2020(3)}. This is a key result of the present work, being an analogous situation found on the verge of the Mott metal-to-insulator transition \cite{Mariano2020(3), MarianoPRL}. It is worth mentioning that, according to the FHS theory, upon considering distinct protein/solvent molecular lengths, the shape of the binodal and spinodal lines can become asymmetric, which in turn affects the value of $\phi_p$ in which $\Gamma$ is enhanced and changes sign. Now, we discuss the dynamics of the protein compartmentalization in the frame of Avramov's/Casalini's model \cite{Avramov1988,Casalini2004}. The latter considers an inherent random spatial disorder, being the volume-dependent relaxation time $\tau$ given by \cite{Casalini2004,Mariano2019(2), Mariano2020(3)}:
\begin{equation}
\tau = \tau_0\exp{\left(\frac{C}{Tv^{\gamma}}\right)},
\label{relaxationtime}
\end{equation}
where $\tau_0$ is a characteristic time-scale, $C$ a non-universal constant \cite{SM}, and $\gamma = v\alpha_{\wp}/\kappa_T c_v$ the effective GP \cite{Mariano2016}, with $v$ the volume, $\kappa_T$ the isothermal compressibility, and $c_v$ the isovolumetric heat capacity. We consider that the random spatial disorder is associated with the presence of droplets in the two-phases region [Fig.\,S1d)]. Indeed, protein droplets are randomly distributed within the cell, giving rise to spatially disordered protein conglomerates \cite{Dyson2015}. The relaxation time is usually defined as the time-scale for a system to reach back its equilibrium after the removal of a perturbation \cite{Mariano2020(3)}. In practice, the relaxation time of living organisms can be accessed, for instance, employing fluorescence recovery after photobleaching measurements \cite{fanbai}. Following discussions in Ref.\,\cite{Colby2003}, we consider $\tau$ as the time-scale required for a synthesized protein to diffuse within the cell. The droplets dynamics is governed by the rate $k_{in}$ ($k_{out}$) in which proteins enter (leave) the droplets, being $k_{in}, k_{out} \propto \tau_0^{-1}$, with $\tau_0 \simeq 10^2$\,s a typical time scale for biological systems \cite{Klosin2020}. In the frame of the FHS theory, it is considered that there are no total volume changes on mixing, so that the volume variation in our case lies solely on the protein molecules added into the system upon increasing $\phi_p$, so that $\kappa_T = -1/V_p(\partial V_p/\partial {\wp})_T$, where $V_p = \phi_pV_{tot}$ is the volume associated with the total number of proteins. Employing basic Thermodynamics \cite{SM}, $\gamma$ can be obtained:
\begin{eqnarray}
&&\gamma = \phi_pV_{tot}[N_p\phi_p + N_s(\phi_p-1)(2\chi N_p\phi_p-1)]\times \nonumber\\
&& \times \{(\phi_p-1)(\phi_pV_{tot}-v_p)[N_p - N_s - \chi N_p N_s + \nonumber \\
&&2\chi N_pN_s\phi_p + N_p\log{(1-\phi_p)}-N_s\log{\phi_p}]\}^{-1},
\label{effectivegamma}
\end{eqnarray}
which is key in our analysis. In Fig.\,\ref{Fig-1}a), $\gamma$ \emph{versus} $\phi_p$ \emph{versus} $T$ is depicted. Note that $\gamma$ is also enhanced and changes sign in the same way as $\Gamma$ \cite{SM}. Employing Eqs.\,\ref{relaxationtime} and \ref{effectivegamma}, $\tau$ can be computed. Note that in the analytical computation of all physical quantities obtained in the present work, we have fixed $0.01 < \phi_p < 0.99$, so that a positivity-preserving constrain is fulfilled.

\section{Discussion}

Remarkably, $\tau \rightarrow \infty$ for values of $\phi_p$ in which both $\Gamma$ and $\gamma$ are anomalous \cite{SM}, namely for $\phi_p = 0.5$ and in the vicinity of the binodal line, cf.\,Fig.\ref{Fig-1}b). This is reminiscent of experimental observations employing static and dynamic light scattering, which demonstrated that on the verge of the binodal line the so-called fast relaxation rates disappear and the system is dominated by slow relaxation rates \cite{woermann}. Such a peculiar behavior of $\tau$ can be interpreted as a slowing down of the protein diffusion dynamics. This is an analogous situation as the ``creation of mass'' observed on the verge of critical points \cite{prbletters}. Note that the dynamics of phase separation is usually obtained employing the Cahn-Hilliard equation given by $F_{CH} = \int[f(\phi_p) + \kappa(\nabla\phi_p)^2]dv$ \cite{rey,zhang}, where $f(\phi_p)$ is the free energy density of a homogeneous system and $\kappa$ a positive constant. The so-called surface diffusion energy term $(\nabla\phi_p)^2$ accounts for diffusion processes associated with phase separation. Interestingly enough, such a term is reminiscent of the phase bending energy $(\nabla\phi^*)^2$ in the frame of generalized rigidity and the famous Higgs mechanism \cite{coleman}, where $\phi^*$ is the phase. Recently, we have proposed the concept of Higgs-like stiffness, in analogy to the Higgs mechanism \cite{hls}. The latter can be extended to any complex physical quantity. In the case of protein compartmentalization, the refractive index might be the suited physical quantity, since it is reported that on the verge of phase separation, enhanced refractive index fluctuations take place, being linked, for instance, to the development of cataracts \cite{foffi}. Hence, although our proposal to compute the protein diffusion time in the frame of a CGLP does not explicitly consider a surface diffusion term, it properly extracts the dynamics in both metastable and unstable regimes of phase separation. Such a divergent-like behavior of $\tau$ minimizes the so-called noise strength $NS$ given by the coefficient of variation squared $CV^2$ [Fig.\,\ref{Fig-1}c)] \cite{SM}. Upon continuously increasing $\phi_p$ from the two phases region at a fixed $T$, first the metastable and, eventually, the single phase region is achieved, cf.\,Fig.\,S1d). Following Landau and Lifshitz's discussions, metastability can be inferred by a negative $\wp$ \cite{landau,negpressure}, giving rise to the spontaneous formation of cavities, in the present case, protein droplets \cite{SM}. We now analyze $NS$ \cite{Klosin2020} in terms of $\tau$ \cite{SM}. Figure\,\ref{Fig-1}c) depicts $CV^2$ as a function of $T$ and $\phi_p$. Note that $CV^2$ is minimized upon approaching the binodal line and $CV^2 \rightarrow 0$ for $\phi_p$ = 0.5, being that $\tau \rightarrow \infty$ in these regimes. The minimization of $NS$ implies that stochastic fluctuations associated with $\Delta F$ are dramatically reduced \cite{Klosin2020}. Such a reduction, together with the enhancement of $\tau$, demonstrates the slow-dynamics, which can be also analyzed in terms of the average protein displacement $\lambda_x$ in the frame of a Brownian motion \cite{SM}, cf.\,Fig.\,\ref{Fig-1}d). Note that $\lambda_x \rightarrow 0$ upon crossing the binodal line and for $\phi_p$ = 0.5, being the CGLP more robust to $\phi_p$ fluctuations around $\phi_p$ = 0.5 given the broad range of the unstable regime observed for $\Delta F$ [Fig.\,S1c)]. This is in line with the fact that $CV^2 \rightarrow 0$ only for $\phi_p = 0.5$. Next, we discuss our findings under the light of the CGLP. Upon reaching the unstable equilibrium condition, i.e., at $\phi_p$ = 0.5, the entropy of mixture $\Delta S$ is maximized, $\gamma \rightarrow \infty$ and, as a consequence, $\tau$ is dramatically enhanced. Upon approaching the binodal line, where the two-phases region is established [Fig.\,\ref{Fig-1}c)], both $\gamma$ and $\tau \rightarrow \infty$, which corroborates our proposal of a CGLP in the vicinity of the binodal line, as well as for $\phi_p = 0.5$. This is because $\wp = 0$ at the binodal line and for $\phi_p$ = 0.5, corroborating our proposal of a slowing-down of the cell dynamics for such particular values of $\phi_p$ \cite{SM}. Considering the two key-ingredients for the establishment of a Griffiths phase \cite{Vojta2013}, namely intrinsic random spatial disorder associated with the protein droplets distribution \cite{Dyson2015} and $\tau \rightarrow \infty$ for particular values of $\phi_p$, cf.\,Fig.\,\ref{Fig-1}b), we introduce unprecedentedly the concept of CGLP. It is worth mentioning that, upon varying $N_p$ and $N_s$, the concentration $\phi_p$ in which the CGLP sets is simply shifted as a consequence of the change of the maximum of $\Delta S$ [Fig.\,S1e)], leading to the \emph{distortion} of the $T$ \emph{versus} $\phi_p$ phase diagram \cite{Colby2003}. The concept of the CGLP can be also extended to other mixtures with more than two components \cite{Venema}. As pointed out in Ref.\,\cite{Munoz2013}, Griffiths-like phases play an important role in bringing local order to globally disordered systems, which is key in improving the functionality of biological systems, such as self-organization and the mechanisms for adaptation and evolution itself. Also, the here-proposed CGLP is a possible explanation to the so-called slow relaxation mode, being its underlying origin still under debate in the literature \cite{chiwu}. Thus, our work paves the way to understand compartmentalization in terms of a CGLP, which can be linked with the origin of primary organisms since only the coacervates with slow-dynamics survived and evolved, which in turn might be related to the key-role played by homochirality in the evolution process of life \cite{SM}.  The enhancement of the protein diffusion time occurs concomitantly with the reduction of the $NS$, which in turn is key in optimizing gene expression \cite{Klosin2020}. In summary, we provide an alternative approach to investigate the dynamics of protein compartmentalization, which is applicable to other biological systems \cite{SM} and can be extended to nematics \cite{SM}. It is challenging to understand the impact of the here-proposed CGLP in the establishment and temporal evolution of various biological processes regarding the impact of LLPS on the treatment of various diseases \cite{Banani2017,SM}. Our analysis is universal and can be extended taking into account, for instance, surface tension and a concentration-dependent $\chi$ \cite{wolf,halperin}.\\

\section{Supplemental information}

Supplemental information can be found online at [link].

\section{Acknowldegements}

MdeS acknowledges financial support from the S\~ao Paulo Research Foundation - Fapesp (Grants No.\,2011/22050-4, 2017/07845-7, and 2019/24696-0), National Council of Technological and Scientific Development - CNPq (Grant No.\,303772/2023-9). ACS acknowledges CNPq (Grant No.\,08695/2021-6). LSR acknowledges the Icelandic Research Fund (Rannis) (Grant No.\,163082-051). This work was partially granted by Coordena\c c\~ao de Aperfei\c coa-\linebreak mento de Pessoal de N\'ivel Superior - Brazil (Capes) - Finance Code 001 (Ph.D. fellowship of IFM). LS acknowledges IGCE for the post-doc fellowship.

\section{Authors contributions}

LS carried out the calculations and generated the figures. LS and MdeS wrote the paper with contributions from IFM, LSR, MFM, AMU, ACS, and RELM. All authors revised the manuscript. MdeS conceived and supervised the whole project.

\section{Declaration of interests}

The authors declare no competing interests.

\section{Data availability}

No data associated with this study was deposited into a publicly available repository. Data included in article/supplementary material/referenced in article.


\begin{thebibliography}{10}
\bibitem{Morgan1927} T.H. Morgan, Science \textbf{65}, 213 (1927).
\bibitem{Schrodinger1944} E. Schr\"odinger, \emph{What is life? The Physical Aspect of the Living Cell} (Cambridge University Press, Cambridge, 1944).
\bibitem{Dronamraju1999} K.R. Dronamraju, Genetics \textbf{153}, 1071 (1999).
\bibitem{Klosin2020} A. Klosin, F. Oltsch, T. Harmon, A. Honingmann, F. J\"ulicher, A.A. Hyman, C. Zechner, Science \textbf{367}, 464 (2020).
\bibitem{Riback2020} J.A. Riback, L. Zhu, M.C. Ferrolino, M. Tolbert, D.N. Mitrea, D.W. Sanders, M.-T. Wei, R.W. Kriwacki, C.P. Brangwynne, Nature \textbf{581}, 209 (2020).
\bibitem{MacMicking2021} S. Huang, S. Zhu, P. Kumar, J.D. MacMicking, Nature \textbf{594}, 424 (2021).
\bibitem{Wang2021}  J.H. Ahn, E.S. Davis, T.A. Daugird, S. Zhao, I.Y. Quiroga, H. Uryu, J. Li, A.J. Storey, Y.-H. Tsai, D.P. Keeley, S.G. Mackintosh, R.D. Edmondson, S.D. Byrum, L. Cai, A.J. Tackett, D. Zheng, W.R. Legant, D.H. Phanstiel, G.G. Wang, Nature \textbf{595}, 591 (2021).
\bibitem{Banani2017} S.F. Banani, H.O. Lee, A.A. Hyman, M.K. Rosen, Nat. Rev. Mol. Cell Biol. \textbf{18}, 285 (2017).
\bibitem{Avramov1988} I. Avramov, A. Milchev, J. Non-Cryst. Solids \textbf{104}, 253 (1988).
\bibitem{Casalini2004} R. Casalini, C.M. Roland, Phys. Rev. E \textbf{69}, 062501 (2004).
\bibitem{Griffiths1969} R.B. Griffiths, Phys. Rev. Lett. \textbf{23}, 17 (1969).
\bibitem{Vojta2013} T. Vojta, AIP Conf. Proc. \textbf{1550}, 188 (2013).
\bibitem{Veis2011} A. Veis, Adv. Colloid Interface Sci. \textbf{167}, 2 (2011).
\bibitem{Mariano2020(3)} I.F. Mello, L. Squillante, G.O. Gomes, A.C. Seridonio, M. de Souza, J. Appl. Phys. \textbf{128}, 225102 (2020).
\bibitem{Miranda2009} E.C. Andrade, E. Miranda, V. Dobrosavljevi\'c, Phys. Rev. Lett. \textbf{102}, 206403 (2009).
\bibitem{Kanoda2020} R. Yamamoto, T. Furukawa, K. Miyagawa, T. Sasaki, K. Kanoda, T. Itou, Phys. Rev. Lett. \textbf{124}, 046404 (2020).
\bibitem{Munoz2013} P. Moretti, M.A. Mu\~noz, Nat. Commun. \textbf{4}, 2521 (2013).
\bibitem{Tragtenberg2016} M. Girardi-Schappo, G.S. Bortolotto, J.J. Gonsalves, L.T. Pinto, M.H.R. Tragtenberg, Sci. Rep. \textbf{6}, 29561 (2016).
\bibitem{Colby2003} M. Rubinstein, R.H. Colby, \emph{Polymer Physics} (Oxford University Press, Oxford, 2003).
\bibitem{debenedetti} P.G. Debenedetti, \emph{Metastable liquids: concepts and principles} (Princeton University Press, Princeton, 1996).
\bibitem{sethna} J.P. Sethna, \emph{Statistical mechanics: entropy, order parameters and complexity} (Oxford University Press, 2021).
\bibitem{overbeek} J.T.G. Overbeek, M.J. Voorn, J. Cell. Comp. Phys. \textbf{49}, 7 (1957).
\bibitem{rpa} V. Yu Borue, I. Ya Erukhimovich, Macromolecules \textbf{21}, 3240 (1988).
\bibitem{stuart} P.M. Biesheuvel, M.A.C. Stuart, Langmuir \textbf{20}, 2785 (2004).
\bibitem{Mariano2016} M. de Souza, P. Menegasso, R. Paupitz, A. Seridonio, R.E. Lagos, Europ. J. Phys. \textbf{37}, 055105 (2016).
\bibitem{Mariano2020} L. Squillante, I.F. Mello, G.O. Gomes, A.C. Seridonio, R.E. Lagos-Monaco, H.E. Stanley, M. de Souza, Sci. Rep. \textbf{10}, 7981 (2020).
\bibitem{MarianoPRL} L. Bartosch, M. de Souza, M. Lang, Phys. Rev. Lett. \textbf{104}, 245701 (2010).
\bibitem{Mariano2019(2)} G. Gomes, H.E. Stanley, M. de Souza, Sci. Rep. \textbf{9}, 12006 (2019).
\bibitem{Marianoarxiv} L. Squillante, I.F. Mello, A.C. Seridonio, M. de Souza, Mater. Res. Bull. \textbf{142}, 111413 (2021).
\bibitem{Mariano2021} L. Squillante, I.F. Mello, A.C. Seridonio, M. de Souza, Sci. Rep. \textbf{11}, 9431 (2021).
\bibitem{Zhu} L. Zhu, M. Garst, A. Rosch, Q. Si, Phys. Rev. Lett. \textbf{91}, 066404 (2003).
\bibitem{SM} See Supplemental Material at [URL] for more information about the computation of $\Gamma$, $\gamma$, and $\tau$, as well as the here-proposed connection between Griffiths phases and Oparin's proposal, nematicity, liquid-liquid phase separation, phenotypes, and diseases treatment.
\bibitem{Dyson2015} P.E. Wright, H.J. Dyson, Nat. Rev. Mol. Cell Bio. \textbf{16}, 18 (2015).
\bibitem{fanbai} X. Jin, J.-E. Lee, C. Schaefer, X. Luo, A.J.M. Wollman, A.L. Payne-Dwyer, T. Tian, X. Zhang, X. Chen, Y. Li, T.C.B. McLeish, M.C. Leake, F. Bai, Sci. Adv. \textbf{7} eabh2929 (2021).
\bibitem{prbletters} L. Squillante, L.S. Ricco, A.M. Ukpong, R.E. Lagos-Monaco, A.C. Seridonio, M. de Souza, Phys. Rev. B. \textbf{108}, L140403 (2023).
\bibitem{woermann} A. Ritzl, L. Belkoura, D. Woermann, Phys. Chem. Chem. Phys. \textbf{1}, 1947 (1999).
\bibitem{rey} P.K. Chan, A.D. Rey, Comput. Mater. Sci. \textbf{3}, 377 (1995).
\bibitem{zhang} L. Dong, C. Wang, H. Zhang, Z. Zhang, Commun. Math. Sci. \textbf{17}, 921 (2019).
\bibitem{coleman} P. Coleman, Introduction to many-body Physics (Cambridge University Press, Cambridge, 2016).
\bibitem{hls} L. Squillante, A.C. Seridonio, R.E. Lagos-Monaco, M. de Souza, Higgs-like stiffness and fractons on the verge of phase transitions (submitted).
\bibitem{foffi} N. Dorsaz, G.M. Thurston, A. Stradner, P. Schurtenberger, G. Foffi, Soft Matter \textbf{7}, 1763 (2011).
\bibitem{landau} Lev D. Landau, Evgeny M. Lifshitz, Statistical Physics (vol.5), (Butterworth-Heinmann, 1980).
\bibitem{negpressure} F.F. Barbosa, L. Squillante, R.E. Lagos-Monaco, A.C. Serdionio, M. de Souza (in preparation).
\bibitem{Venema} A. Bot, E. van der Linden, P. Venema, ACS Omega \textbf{9}, 22677 (2024).
\bibitem{chiwu} J. Li, T. Ngai, C. Wu, Polym. J. \textbf{42}, 609 (2010).
\bibitem{wolf} B.A. Wolf, Macromol. Chem. Phys \textbf{204}, 1381 (2003).
\bibitem{halperin} V.A. Baulin, A. Halperin, Macromolecules \textbf{35}, 6432 (2002).
\end{thebibliography}
\end{document}